\documentstyle[11pt,a4]{article}

\newcommand{\gsim}{\stackrel{\lower.7ex\hbox{$>$}}{\lower.7ex\hbox{$\sim$}}}
\newcommand{\lsim}{\stackrel{\lower.7ex\hbox{$<$}}{\lower.7ex\hbox{$\sim$}}}

\newcommand\GeV{\,\mbox{GeV}}

\newcommand\TeV{\,\mbox{TeV}}

\newcommand\fB{{\,\mbox{fb}}}

\newcommand\cm{{\,\mbox{cm}}}
\newcommand\second{{\,\mbox{s}}}
\newcommand\MSbar{$\overline{\mbox{MS}}$}
\newcommand\order{{\cal O}}
\newcommand\mat{{\cal M}}

\newcommand\nn{\nonumber\\}
\renewcommand\Re{\mbox{Re}}

\begin{document}
\thispagestyle{empty}

\setcounter{page}{0}

\begin{flushright}
CERN-TH.6581/92\\
MPI-Ph/91-46\\
LMU-92/06\\
July 1992
\end{flushright}
\vspace*{\fill}
\begin{center}
{\Large\bf QCD and QED Corrections to Higgs Boson Production
in
Charged Current $ep$ Scattering}\\
\vspace{2em}
\large
\begin{tabular}[t]{c}
J. Bl\"umlein$^{a,b,}$\footnotemark[1]\\
G.~J. van Oldenborgh$^b$\\
R. R\"uckl$^{b,c,d}$\\
\\
{$^a$ \it Institut f\"ur Hochenergiephysik,
O--1615 Zeuthen, FRG}\\
{$^b$ \it Sektion Physik der Universit\"at M\"unchen,
W--8000 M\"unchen 2, FRG}\\
{$^c$ \it Max-Planck-Institut f\"ur Physik,
Werner-Heisenberg-Institut,}\\
{\it W--8000 M\"unchen 40, FRG}\\
{$^d$ \it CERN, CH-1211 Gen\`eve 23, Switzerland}
\end{tabular}
\end{center}
\vspace*{\fill}

\begin{abstract} First order QCD and leading QED corrections to Higgs boson
production in the channel $e^-p \to \nu H^0 X; H^0 \to b\bar{b}$ are calculated
for the kinematical conditions at LEP $\otimes$ LHC ($\sqrt{s} = 1360 \GeV$)
and the interesting mass range $80 < M_H < 150 \GeV$. In the DIS scheme the QCD
corrections (not including the corrections to the branching ratio, which are
well-known) are found to be about 1\% for the total cross section and $-13\%$
to $-10\%$ for the observable cross section as defined by appropriate cuts.
The latter results depend on the definition of these cuts. The QED corrections
amount to about $-5\%$.

\end{abstract}
\vspace*{\fill}

{\small\noindent\footnotemark[1] This work was supported in parts by the
Alexander-von-Humboldt Stiftung and German BMFT under contract number
055ZT91I--01

\newpage


\section{Introduction}
\label{sec-int}

The search for the Higgs boson, or physics which replaces the Higgs boson in
case it does not exist as a fundamental particle, is one of the main
motivations for the construction of supercolliders. While a light Higgs boson
with a mass $M_H \leq M_W$ will most likely be found or excluded at LEP, the
mass range from $M_H \approx 140 \GeV$ up to about $1 \TeV$ can be completely
covered at LHC and SSC. However, the intermediate region from $M_H \approx 80
\GeV$ to $140 \GeV$ is a very difficult one for searches in $pp$ collisions
\cite{PPCOL}. In this mass range, the Higgs boson decays dominantly into $b
\overline{b}$ quark pairs, a channel which is difficult to isolate because of
the huge QCD background. The rare decay $H^0 \to \gamma \gamma$, on the other
hand, provides a more favourable signature, but at the cost of a substantial
reduction in rate \cite{PPCOL,XX}. Although it appears feasible to detect some
signal in the two photon channel, this task is experimentally very demanding.
Clearly, the best tool for the discovery of an intermediate mass Higgs boson
would be a linear $e^+e^-$ collider at a minimum cms energy of roughly twice
the LEP 200 energy \cite{JLC}. Unfortunately, at the time when LHC and SSC are
expected to start operation, a linear collider in the required energy range
will most likely not yet be available. Therefore, it is an important issue to
clarify whether or not the $ep$ option which exists at LHC could help to
investigate this very interesting mass window.

The $ep$ option we refer to can be realized by intersecting a $50$ to $100
\GeV$ electron beam from LEP with the $7.7 \TeV$ proton beam from LHC
\cite{LUM}. In this mode, the expected luminosity ranges from $2 \cdot 10^{31}
\cm^{-2}\second^{-1}$ at the upper end of the cms energy range, $\sqrt{s} =
1.79 \TeV$, to $2 \cdot 10^{32} \cm^{-2}\second^{-1}$ at the lower end,
$\sqrt{s} = 1.26 \TeV$. For definiteness, we assume $\sqrt{s} = 1.36 \TeV$ and
an integrated luminosity of $1 \fB^{-1}$ corresponding to collisions of $60
\GeV$ electrons on $7.7 \TeV$ protons and one year of running time.

The basic electroweak processes in $ep$ collisions are neutral current (NC) and
charged current (CC) scattering through the exchange of virtual photons, $Z$
and $W$ bosons, respectively. Whereas the couplings of the standard Higgs
scalar to the fermions participating in NC and CC scattering are strongly
suppressed by the light fermion masses, the couplings to the $W$ and $Z$ bosons
are of ordinary electromagnetic strength. Hence, the virtual weak bosons
appearing in NC and CC scattering processes can act as efficient sources for
the standard Higgs boson. In fact, the $WW$ fusion process depicted in
fig.~\ref{F1} is the dominant Higgs production mechanism in $ep$ collisions
\cite{BORN1,BORN2}, followed by $ZZ$ fusion, which has a roughly 5 times
smaller cross section. Interestingly, in $pp$ collisions the analogous
mechanisms become important only for very heavy Higgs bosons with masses $M_H
\geq 600 \GeV$ \cite{PPCOL}.

As can be already expected from the above remarks, Higgs bosons are not
produced very frequently in $ep$ collisions, at least in comparison to the
corresponding rates in $pp$ collisions. In the interesting mass range from 80
to 140 GeV and for $\sqrt{s} = 1.36 \TeV$, the total cross section varies from
about 200 to 100 fb. One can therefore not afford to search in rare decay
channels such as $H^0 \to \gamma \gamma$, the mode considered for hadron
colliders, but one must try to detect the Higgs signal in the main decay
channel $H^0 \to b \overline b$ (or for $M_H \geq 130 \GeV$ through $H^0 \to W
W^*$, where one of the $W$ bosons is off-shell, 
this possibility is not pursued further here). The main
background is expected to come from NC production of jets (including
photoproduction), multi-jet CC scattering, and single $W\!$, $Z$ and $t-$quark
production with subsequent hadronic decays. Although this background is large,
it is not as overwhelming as the QCD jet background to $H^0 \to b
\overline{b}$ in $pp$ collisions. In addition, in $ep$ collisions the $H^0 \to
b \overline{b}$ decay provides several characteristic signatures which allow
for an efficient discrimination of signal against background. For the $WW$
fusion channel $ep \to \nu H^0 q + X \to \nu b \overline{b} q + X$, the
selection of signal events and the suppression of the backgrounds have been
investigated in great detail in ref.~\cite{BORN3}. It has been demonstrated
that one should be able to observe an  intermediate mass Higgs boson, provided
flavour identification capabilities are available with efficiencies similar to
the expected capabilities of the DELPHI detector at LEP \cite{DELPHI}.

This encouraging result and the importance of the physics issue call for
further studies in order to corroborate the above conclusion. Obviously, the
feasibility of flavour identification has to be examined in a dedicated $ep$
detector study. Furthermore, one has not yet exploited all possibilities to
optimize the event selection and background suppression. Finally, the parton
level treatment in tree approximation performed in ref.~\cite{BORN3} still
suffers from uncertainties due to the neglect of hadronization effects and
higher order corrections. In particular, K-factors of 2 would be quite crucial
since the estimated number of observable signal events and the signal to
background ratio do not leave much room for losses. Therefore, it is necessary
to investigate the effects of higher order corrections to signal and
background. One cannot rely on qualitative arguments and guesses based on
existing calculations for production processes of other heavy particles
\cite{CCQCD} since the Higgs signal considered here is defined by a set of
nontrivial cuts on several kinematic variables.  In this paper, we calculate
the $\order(\alpha_s)$ QCD and leading QED corrections to the $WW$ fusion
process in the zero-width approximation for the Higgs boson, both for the fully
integrated cross section as well as for the observable cross section in the
channel $H^0\to b\bar{b}$ after application of the kinematical cuts suggested
in ref.~\cite{BORN3}.

The paper is organized as follows. In section~\ref{sec-born} we recalculate the
tree level cross section for Higgs production via $WW$ fusion and examine the
variation with different sets of structure functions. We also show interesting
differential distributions for the channel $H^0 \to b \overline{b}$ and
briefly review the selection cuts used in ref.~\cite{BORN3}.
Section~\ref{sec-QCD} is devoted to the QCD corrections. Results are presented
for the fully integrated production cross section, including the QCD corrected
branching ratio for $H^0 \to b \overline{b}$. The main point, however, is the
calculation of the QCD corrections to the cross section times branching ratio
in the three jet channel in the presence of the kinematical cuts of ref.\
\cite{BORN3}. The effects of initial state QED bremsstrahlung are then
obtained in section~\ref{sec-QED} using the structure function method. In
section~\ref{sec-concl} we summarize our conclusions.


\section{The Born Approximation}
\label{sec-born}

The lowest order production cross section for $e^-p \rightarrow \nu H^0 X$ is
obtained from the diagram shown in fig.~\ref{F1}.  The differential cross
section is given by\footnote{For
$e^+p$-scattering the distributions $u_i, \bar{d}_i$ have to be replaced by
$d_i,\bar{u}_i$.}
\begin{eqnarray}
\label{eq1}
    d\sigma^{(0)} & = & \frac{1}{2xs} \sum_i |\mat_i^{(0)}|^2
    q_i(x,Q_2^2)\,dP\!S_3(e q\to \nu H^0 q)\,dx
\nn
    & = & \frac{1}{2xs} \frac{g^6 M_W^2}
    {(Q_1^2 + M_W^2)^2 (Q_2^2 + M_W^2)^2}
    \sum_{i=1}^2 \Bigl[(p_1 p_2)(p'_1 p'_2) u_{i}(x,Q^2_2)
\nn&&\mbox{}\quad\quad
    + (p_1 p'_2)(p'_1 p_2) \bar{d}_{i}(x,Q^2_2) \Bigr]
    \,dP\!S_3(e q\to \nu H^0 q)\,dx
\end{eqnarray}
where the matrix element squared includes an average of the spins of incoming
particles and a sum over the spins of outgoing particles.  In the above
formula $p_1$($p'_1$) and $p_2$($p'_2$) are the incoming (outgoing) lepton and
quark momenta, respectively, $x$ is the fraction of the proton momentum carried
by the incoming partons, $s$ the cms energy squared and $Q_i^2 =-(p_i -
p'_i)^2$.  $u_i(x,Q^2)$ are the distribution functions of $u-$ and $c-$quarks,
$\bar{d}_i(x,Q^2)$ of $\bar{d}-$ and $\bar{s}-$ antiquarks. The contributions
{}from bottom and top quark densities can be neglected.  Also we have put
off-diagonal elements of the CKM-matrix involving the top quark to zero.  The
SU(2) gauge coupling $g$ is given by $g = e/\sin \theta_W$, $e$ being the
electromagnetic coupling constant and $\theta_W$ being the weak mixing angle.
Furthermore, we apply the zero-width approximation, which is a valid
simplification for $M_H \leq 150 \GeV$ where $\Gamma_H/M_H \lsim 10^{-4}$
\cite{BRANCH1,BRANCH1A}.  The phase space element $dP\!S_n$ is given by
\begin{equation}
    dP\!S_n = (2\pi)^{(4-3n)}\,
    \delta^{(4)}(p_{\rm in}-\sum_{i=1}^n p_i)
    \prod_{i=1}^n \frac{d^3 p_i}{2p_{i,0}}.
\end{equation}
The physical boundaries in $x$ are $M_H^2/s \leq x \leq 1$.
The integrated cross section $\sigma^{(0)}_{tot}$  was checked against
earlier calculations \cite{BORN2,BORN3}. Throughout this paper we have taken
the following values for the electroweak parameters: $M_W = 80.6 \GeV$, $\sin^2
\theta_W = 0.230$, and $e^2/4\pi = 1/128.5$. The quark distributions in eq.
\ref{eq1} are parametrized as $Q^2$ dependent functions corresponding to the
solutions of the Altarelli-Parisi equations.

In fig.~\ref{F2} we show $\sigma^{(0)}_{tot}$ and $\sigma^{(0)}_{tot} \times
Br(H^0\to b\bar{b})$.  The branching ratio is taken from ref.~\cite{BRANCH1A}
and includes integrated QCD corrections up to $\order(\alpha_s^2)$.  The fast
decrease of the cross section times branching ratio in the $b\bar{b}$ channel
at higher Higgs boson masses is due to the decay $H^0\to WW^*$ which becomes
dominant as $M_H$ exceeds 120 \GeV.

Concerning the dependence of the cross section on the parametrization of the
quark distribution functions the following remarks may suffice. We use the
parametrization by Morfin and Tung~\cite{MT} set~2 assuming the \MSbar\ or DIS
scheme  \cite{DIS} for the definition of the parton densities. The resulting
$\sigma_{tot}^{(0)}$ for the two schemes differ by less than 3\%, as can be
seen from fig.~\ref{F2}. We have also compared $\sigma^{(0)}_{tot}$ for  other
parton parametrizations. The deviations for HMRS (\MSbar, set 2) \cite{HMRS}
are also smaller than 3\%. Larger differences of $\order(10\%)$ are obtained
for older parametrizations such as DFLM (DIS) \cite{DFLM} and DO1 (LO)
\cite{DO}.

Eq.~\ref{eq1} was parametrized in terms of the individual parton densities
directly.  However, for the understanding of the size of the QCD corrections to
be discussed in the next section, it is interesting to know which of the usual
deep inelastic structure functions, that is
\begin{equation}
\label{F2eq}
2 x F_1^{(0)}(x,Q^2_2) =
F_2^{(0)}(x,Q^2_2) = 2 x \sum_{i=1}^2 [ u_i(x,Q^2_2) +
\bar{d}_i(x,Q^2_2)]
\end{equation}
and
\begin{equation}
\label{F3eq}
x F_3^{(0)}(x,Q^2_2) = 2 x \sum_{i=1}^2 [ u_i(x,Q^2_2) -
\bar{d}_i(x,Q^2_2)],
\end{equation}
yields the dominant part of the cross section.  For
this purpose, eq.~\ref{eq1} may be rewritten as
\begin{equation}
\label{eq1a}
d\sigma^{(0)} = \frac{1}{2xs} \frac{g^6 M_W^2}
{(Q_1^2 + M_W^2)^2 (Q_2^2 + M_W^2)^2}
\left \{ b^+ F_2^{(0)}(x,Q^2_2) / x  + b^-
F_3^{(0)}(x,Q^2_2) \right \}\,dP\!S_3\,dx
\end{equation}
with
\begin{equation}
\label{eq:bplusmin}
b^{\pm} = \frac{1}{4} [(p_1 p_2)(p'_1 p'_2)
\pm (p_1 p'_2)(p'_1 p_2)].
\end{equation}
$F_2^{(0)}(x,Q^2)$ contributes about 98\% to $\sigma^{(0)}_{tot}$. This has
two reasons: $b^- \ll b^+$ in a wide range of the phase space\footnote{For
$p_H \rightarrow 0$, this corresponds to deep inelastic scattering at small
$y$.}, and $xF_3^{(0)} \leq F_2^{(0)}$.

To illustrate some details of the $WW-$fusion process differential
distributions for $x = -q^2_2/2(Pq_2)$ ($P$~being the proton momentum),
$Q^2_2~=~-q^2_2$ and the transverse momentum $p_\perp$ and rapidity $\eta$ of
the Higgs boson are shown in fig.~\ref{F3}.  The figures refer to a Higgs
boson mass of 100~\GeV.  As one can see, the production involves large $x$ and
$Q^2_2$, typically $x \gsim 0.01$ and $Q^2_2 \gsim 100 \GeV^2\ $. We thus do
not expect screening corrections to the parton distributions \cite{SCREEN} to
influence the cross section significantly.\footnote{We verified this using the
KMRS distributions (\MSbar, B-5) \cite{KMRS}.}

The background to the $\nu b \bar{b} X$ final state was investigated in
ref.~\cite{BORN3} in detail. It receives contributions from four types of
reactions: photo(and NC)production of multijet events, CC multijet production,
single $W$ and $Z$ production with subsequent hadronic decay, and top
production.  This background was shown to be manageable after applying the
following selection criteria:
\begin{itemize}
\item the final state should contain three jets, defined
here as partons with
transverse momentum $p_\perp > 20 \GeV$, rapidity
$|\eta| < 4.5$ and a
distance $\Delta R > 1$ in the $\eta$-$\phi$ plane;
\item the event should have a large missing transverse momentum,
$p_\perp^{miss} > 20 \GeV$, and a large total transverse
energy, $E_\perp > 100
\GeV$;
\item the Higgs signal should be contained in the invariant mass
distribution
of the two jets lowest in rapidity.
\end{itemize}
These cuts leave from the total cross section in the $b\bar{b}$ channel,
$\sigma_{tot}^{(0)}\times Br(H^0\to b\bar{b})$, the fraction denoted by
$\sigma_{obs}^{(0)}$ in figure~\ref{F2}.  In this sense, $\sigma_{obs}^{(0)}$
can be considered the observable cross section in this channel. If one
furthermore assumes good flavour identification and a 10 GeV mass resolution
the background is reduced to a level which would enable detection of the Higgs
boson between 80 and 140 GeV \cite{BORN3}.


\section{QCD Corrections}
\label{sec-QCD}


The QCD corrections to $e^- p \rightarrow \nu H^0 X; H^0 \rightarrow b\bar{b}$
consist of three contributions: the corrections to the W-quark vertex $W^*q\to
q'$, corrections to the decay vertex $H^0 \rightarrow b \bar{b}$, and gluon
exchange between the final state $b$-quarks and the interacting parton at the
hadronic side.  To $\order(\alpha_s)$ the latter corrections vanish because of
the colour structure.

We first consider the total integrated corrections.  They only need to be
calculated for the production vertex, as the integrated corrections to the
decay are known \cite{BRANCH1,BRANCH1A} and have already been taken into
account in the branching ratio.  Then we discuss the QCD corrections to the
observable cross section, which strongly depend on the cuts applied to isolate
the signal.  In this case there is also a negative correction
                                                      to the decay width.


\subsection{Corrections to the total cross section}
\label{sec-QCD1}

The $\order(\alpha_s$) QCD corrections to the total production cross section
are described by the diagrams of fig.~\ref{F5}.  Generally, the differential
cross section for $ep\to\nu H^0 X$ can be written as
\begin{equation}
\label{eqmat}
d\sigma = \frac{1}{2xs} \frac{g^6 M_W^2}{(Q_1^2 + M_W^2)^2 (Q_2^2 + M_W^2)^2}
\left \{
L^{\mu \nu}_{pc}
( W^{pc,L}_{\mu
    \nu} + W^{pc,2}_{\mu \nu}) + L^{\mu \nu}_{pv} W^{pv,3}_{\mu \nu}
\right \}
    \,dP\!S_3\,dx
\end{equation}
where $pc$ and $pv$ denote the parity conserving and violating parts of the
leptonic ($L^{\mu\nu}$) and hadronic ($W_{\mu\nu}^j$) tensors, respectively.
The hadronic tensors $W^j_{\mu \nu}$ are the same as those which appear in
deep inelastic charged current scattering with $Q^2 = -q_2^2$
(cf.~\cite{STRUC1,STRUC2,STRUC3}). For the individual contributions to
eq.~\ref{eqmat} one obtains\footnote{In the limit $p_H \rightarrow 0$,
eqs~\ref{eqF1}--\ref{eqFF3} yield the familiar expressions for the deep
inelastic scattering cross section: $L^{\mu \nu}_{pc} W^{pc,L}_{\mu \nu}
\rightarrow -(sx/4)^2 y^2 {\cal F}_L(x,Q^2)/x$, $L^{\mu \nu}_{pc} W^{pc,
2}_{\mu \nu} \rightarrow (sx/4)^2 Y_+ {\cal F}_2(x,Q^2)/x$ and $L^{\mu
\nu}_{pv} W^{pv,3}_{\mu \nu} \rightarrow (sx/4)^2 Y_- {\cal F}_3(x,Q^2)$, with
$Q^2 = Q^2_1 = Q^2_2$ and $Y_{\pm} = 1 \pm (1-y)^2$.}
\begin{eqnarray}
\label{eqF1}
L^{\mu \nu}_{pc} W^{pc,L}_{\mu \nu} & = & b^L {\cal F}_L(x,Q_2^2)/x
\\
\label{eqF2}
L^{\mu \nu}_{pc} W^{pc,2}_{\mu \nu} & = & b^+ {\cal F}_2(x,Q^2_2)/x
\\
\label{eqFF3}
L^{\mu \nu}_{pv} W^{pv,3}_{\mu \nu} & = & b^- {\cal F}_3(x,Q^2_2).
\end{eqnarray}
Formally, $b^\pm$ can be taken from eq.~\ref{eq:bplusmin} with $p_2=xP$,
$p_2'=p_2+q_2$ not necessarily corresponding to particle momenta (only at
lowest order).  The coefficient $b^L$ is given by
\begin{equation}
b^L(p_1,p'_1,p_2,p'_2)  =  \frac{1}{8} \left \{
(p_1 q_2) (p_1' q_2) - (p_1 p_1') (p_2 q_2) \right\}.
\end{equation}
Here, the structure functions are defined by
\begin{equation}
{\cal F}_i(x,Q^2) = F_i^{(0)}(x,Q^2) + \frac{\alpha_s(Q^2)}{2\pi}
F_i^{(1)}(x,Q^2) + \order(\alpha_s^2)
\end{equation}
with $F_L^{(0)}(x,Q^2) = 0$ and $F_{2,3}^{(0)}(x,Q^2)$ as given in eqs
(\ref{F2eq}) and (\ref{F3eq}).  For $\alpha_s(Q^2)$ the leading order
expression for five active flavours is used.  The value of the QCD scale
$\Lambda$ is chosen in accordance with the parton distribution
functions.\footnote{Since we are calculating the finite terms of the order
$\alpha_s$ corrections, next to leading order parton densities and
$\alpha_s(Q^2)$ should be used.  However, as the final corrections turn out to
be small the difference is negligible.}

For the regularisation of the collinear divergences a factorisation scheme has
to be chosen.  Because we consider a deep inelastic process the choice of the
DIS scheme \cite{DIS} appears to be most natural.\footnote{Note that for
higher order corrections one can choose the schemes to renormalize the $\beta$
function and the collinear divergences separately.}  Different corrections
would be obtained using other schemes, e.g.\ the \MSbar\ scheme.

In the DIS scheme the structure function
\begin{equation}
\label{DISF2}
{\cal F}_2(x,Q^2) = F_2^{(0)}(x,Q^2)
\end{equation}
is preserved
to all orders and corrections occur to the terms
$L^{\mu \nu}_{pc} W^{pc,L}_{\mu \nu}$
and
$L^{\mu \nu}_{pv} W^{pv,3}_{\mu \nu}$ only. In lowest order,
eqs~\ref{eqmat}--\ref{eqFF3} are seen to lead to
$d\sigma^{(0)}$ as given in eq.~\ref{eq1a}.
The QCD diagrams of fig.~\ref{F5},
integrated over all phase space, yield
\begin{eqnarray}
\label{MCAL}
\sigma^{(1)}(QCD) & = & \int\!\!\!dx \int\!\!dP\!S_3 \,
\frac{1}{2xs}
\frac{g^6 M_W^2}
{(Q_1^2 + M^2_W)^2 (Q_2^2 + M^2_W)^2}
\frac{\alpha_s(Q^2_2)}{2 \pi}
\nn&&\quad\quad\mbox{}
\times
\left \{ b^L F_L^{(1)}(x,Q^2_2)/x
            + b^- F_3^{(1)}(x,Q^2_2) \right \}
\end{eqnarray}
where
\begin{eqnarray}
\label{QFL1}
F_L^{(1)}(x,Q^2) & = &
x \int^1_x \frac{dz}{z} \left \{ f_L^q(z) \frac{z}{x}
F_2^{(0)}(x/z,Q^2)
+ 4 f_L^G(z) G(x/z,Q^2) \right \} \\
\label{QFL3}
F_3^{(1)}(x,Q^2) & = &
- C_F \int^1_x \frac{dz}{z} (1+z) F_3^{(0)}(x/z,Q^2).
\end{eqnarray}
Here, $f_L^q(z) = 2 C_F z$ and $f_L^G(z) = 8 T_R z (1-z)$,
with $C_F = 4/3$ and $T_R = 1/2$.
$G(x,Q^2)$ denotes the gluon density. The scale in both
the distribution
functions and $\alpha_s$ is taken to be $Q^2 = -q_2^2$,
in accordance with the
definition of the DIS scheme.

The QCD corrections to the total cross section obtained from eq.~\ref{MCAL}
are displayed in fig.~\ref{F6}.  We show separately the quark and gluon
contributions from $F^{(1)}_L$, $\sigma_L^q$ and $\sigma_L^G$, and the
contribution $\sigma_3^q$ from $F_3^{(1)}$. In the mass range $50 < M_H < 150
\GeV$ the correction is dominated by the $F_L^{(1)}$ term, to which quarks and
gluons contribute about equally. The $\sigma_3^{(1)}$ correction is negative
and amounts to less than 10\% of $\sigma_L^q$.  The total correction
$\sigma^{(1)}_{tot}(QCD)$ adds up to less than 1\% of the Born cross section.
Using for the parametrization of the parton densities DO1 \cite{DO} instead of
MT \cite{MT} the relative correction $\delta^{(1)}_{tot}(QCD) =
\sigma^{(1)}_{tot}(QCD)/\sigma^{(0)}_{tot}$ does not change significantly.
This reflects the similar shape of the parton densities in the region in $(x,
Q^2)$ which contributes most to the cross section.

As a consequence of the above result we remark
that the integral $\order(\alpha_s)$ correction to
Higgs production in $pp$ collisions proceeding via $WW$ fusion is of
$\order(2 \%)$.


\subsection{Inclusion of kinematical cuts}
\label{sec-QCD2}

\subsubsection{Corrections to the production cross section}
\label{sec-QCD21}

A Higgs signal in the $b\bar{b}$ channel can only be observed after applying
suitable cuts (see section~\ref{sec-born}). These cuts strongly influence the
QCD corrections, as exactly three observable jets are demanded.  This means
that there is an upper bound on the angle and energy of the extra gluon or
quark with respect to the other three jets.  These bounds can only be
implemented numerically.  We have chosen to regulate the divergences as
follows: the ultra violet divergences using the usual dimensional
regularization, the infra red divergences with a small gluon mass $\lambda$,
and the collinear divergences (mass singularities) with a small quark mass $m
\gg \lambda$.  This way all phase space integrals can be evaluated numerically
in four dimensions.  This mass regularisation scheme has been described in
ref.\ \cite{MASS}.

The  total corrections  can then  be written  as a  sum of  four 4-dimensional
integrals:
\begin{equation}
\sigma^{(1)}(QCD) = \sigma^{(1)}_{virt+soft} + \sigma^{(1)}_{hard} +
    \sigma^{(1)}_{glue} + \sigma^{(1)}_{counter}.
\end{equation}
The first two  originate in virtual  and real gluon  radiation.
$\sigma^{(1)}_{glue}$ is  the contribution  from initial  state gluons, while
the counter terms arise from the renormalization of the distribution functions
and are needed to cancel the collinear divergences.

The virtual and soft contribution can be written in the form
\begin{eqnarray}
\label{eq:matvirtsoft}
d\sigma^{(1)}_{virt+soft} & = & \frac{1}{2xs}\ \sum_{i=1}^2 \left\{
    2 \Re \Bigl(\mat_{i,virt}^{(1)} \mat_i^{(0)\dagger} \Bigr)
    + |\mat^{(1)}_{i,soft}|^2 \right\} q_i(x,Q_2^2)
    \,dP\!S_3 \,dx
\nn
    & = & \frac{1}{2xs}\frac{g^6 M_W^2}
    {(Q_1^2 + M_W^2)^2 (Q_2^2 + M_W^2)^2}
    \sum_{i=1}^2 \left  [(p_1 p_2)(p'_1 p'_2) u_{i,virt+soft}^{(1)}(x,Q^2_2)
    \right.
\nn&&\quad\quad\quad\quad
    \left. + (p_1 p'_2)(p'_1 p_2) \bar{d}_{i,virt+soft}^{(1)}(x,Q^2_2)
    \right] \,dP\!S_3 \,dx
\end{eqnarray}
where the functions $q_{i,virt+soft}^{(1)}$ are given in the appendix,
eq.~(\ref{eq:qvirtsoft}).  They still depend on the quark mass $m$ and the
soft cutoff $\Delta = (p_2' + k)^2_{min}$, where $k$ is the gluon momentum.
The expressions for $d\sigma^{(1)}_{hard}$ and $d\sigma^{(1)}_{glue}$ are also
provided in the appendix by eqs~(\ref{eq:hardquark}) and (\ref{eq:matglue}),
respectively.

Because we decided to calculate the QCD corrections in the DIS scheme,
demanding eq.\ (\ref{DISF2}) generates a counter term to the quark
distribution functions:
\begin{equation}
\label{eqcount}
    q_{i,counter}^{(1)}(x,Q^2) = - \left \{
        q_{i,virt}^{(1)}(x,Q^2) +
        q_{i,soft}^{(1)}(x,Q^2) +
        q_{i,hard}^{(1)}(x,Q^2) +
        q_{i,glue}^{(1)}(x,Q^2) \right \}
\end{equation}
The various contributions to (\ref{eqcount}) were derived in several
previous calculations \cite{STRUC1,STRUC2,STRUC3} using different
renormalization methods. They are listed in the appendix (eqs
(\ref{eq:countervir})--(\ref{eq:counterglue})) for the
regularization method used in the present calculation.
The ultra violet and infra red singularities as well as the dependence on the
parameter $\Delta$
separating the soft and hard part of the gluon Bremsstrahlung terms cancel in
the sum eq.~(\ref{eqcount})
and only the mass singularity, that is the logarithmic dependence on $m$,
remains. One obtains
\begin{eqnarray}
\label{eq:counter2}
    q^{(1)}_{i,counter}(x,Q^2) & = &
    - \frac{\alpha_s}{2 \pi}  \int_0^1 \frac{d\xi}{\xi}
    \left \{ \widehat{P}_{qq} \Bigl(\xi, \frac{m^2}{Q^2} \Bigr)
    \Bigl[ \theta(\xi -x) \, q_i \Bigl(\frac{x}{\xi},Q^2 \Bigr)
    - \xi \, q_i(x,Q^2) \Bigr] \right.
\nn&&\quad\quad\quad
    + \left.
    \widehat{P}_{qG}\Bigl(\xi, \frac{m^2}{Q^2} \Bigr) \,
    \theta(\xi -x) \, G\Bigl(\frac{x}{\xi},Q^2 \Bigr)
    \right\}
\end{eqnarray}
where the functions $\widehat{P}_{qq}$ and $\widehat{P}_{qG}$ are given in
eqs~(\ref{eq:Pqq}) and (\ref{eq:PqG}) respectively.  Using eq.\
(\ref{eq:counter2}) one finally has
\begin{eqnarray}
\label{eq:dsigmacount}
d\sigma^{(1)}_{counter} & = & \frac{1}{2xs}\ \sum_{i=1}^2
    {|\mat_i^{(0)}|^2} q^{(1)}_{i,counter}(x,Q_2^2)
    \,dP\!S_3 \,dx
\nn
    & = & \frac{1}{2xs}\frac{g^6 M_W^2}
    {(Q_1^2 + M_W^2)^2 (Q_2^2 + M_W^2)^2}
    \sum_{i=1}^2 \left  [(p_1 p_2)(p'_1 p'_2) u_{i,counter}^{(1)}(x,Q^2_2)
    \right.
\nn&&\quad\quad\quad\quad
    \left. + (p_1 p'_2)(p'_1 p_2) \bar{d}_{i,counter}^{(1)}(x,Q^2_2)
    \right] \,dP\!S_3 \,dx.
\end{eqnarray}

In the practical calculation we combine the counter term given in eq.\
(\ref{eq:dsigmacount}) with the virtual and soft contribution, eq.\
(\ref{eq:matvirtsoft}).  Using eq.\ (\ref{eqcount}) one sees that the virtual
and soft parts cancel, and only the hard and glue part of the counter terms
needs to be evaluated.

It turns out that the computational procedure outlined in ref.~\cite{MASS}
(straight evaluation of the individual integrals using Monte Carlo
techniques) is not suitable here, since the corrections are very small.
The problem is that we still have large logarithms that are subtracted
{\em after} the numerical integration over the phase space.  The unavoidable
inaccuracy in this integration (usually $\order(10^{-3})$) then gives a very
large error in the result.
This problem was avoided by performing the subtraction of the collinear
singularity in the  hard Bremsstrahlung and glue contributions, eqs
\ref{eq:hardquark} and \ref{eq:matglue}, respectively, and in the counter
terms {\em before\/} integrating over the rest of phase space. One starts by
rewriting the order of these integrals:\footnote{This is for the quark
contribution.  The procedure for $\sigma_{glue}^{(1)}$ is analogous.}
\begin{eqnarray}
    \sigma^{(1)}_{hard} & = &
    \!\int_0^1\!\! dz \frac{1}{2zs} \int\!\! dP\!S_4(eq\to\nu H^0 q g) \,
    \theta(\hat{s}-\Delta)
    \sum_{i=1}^2 |\mat^{(1)}_{i,hard}|^2 q_i(z,Q_2^2)
\nn & = &
    \!\int_0^1\!\!dz \frac{1}{2zs} \int\!\! d\hat{s}
    \,\theta(\hat{s}-\Delta)
    \int\!\! dP\!S_3(eq\to\nu H^0 q^*) \frac{1}{2\pi}
    \int\!\!dP\!S_2(q^* \to qg)\,
\nn&&\quad\quad \times
    \sum_{i=1}^2 |\mat^{(1)}_{i,hard}|^2 \,
    q_i(z,Q_2^2)
\\
    \sigma^{(1)}_{counter,hard} & = &
    - \frac{\alpha_s}{2\pi}
    \!\int_0^1\!\! dx \frac{1}{2xs} \!\int_x^{1-\delta}\!\! \frac{d\xi}{\xi}
    \!\int\!\! dP\!S_3(eq\to\nu H^0 q) \, \sum_{i=1}^2 |\mat_i^{(0)}|^2
    \,\widehat{P}_{qq} \Bigl(\xi, \frac{m^2}{Q_2^2} \Bigr)
\nn&&\quad\quad
    \times q_i\Bigl(\frac{x}{\xi},Q_2^2\Bigr)
\nn & = &
    - \frac{\alpha_s}{2\pi}
    \!\int_0^1\!\!dz \frac{1}{2zs} \!\int_0^1\!\! \frac{d\xi}{\xi} \,
    \theta(1-\xi-\delta)
    \!\int\!\!dP\!S_3(eq\to\nu H^0 q) \sum_{i=1}^2 |\mat_i^{(0)}|^2
    \,\widehat{P}_{qq} \Bigl(\xi, \frac{m^2}{Q_2^2} \Bigr)
\nn&&\quad\quad
    \times q_i(z,Q_2^2)
\end{eqnarray}
with $\hat{s} = (p_2'+k)^2$ and $\delta=(\Delta-m^2)/Q_2^2$.  Next, we also
introduce in the counter term a gluon momentum $k=p_2(1-\xi)/\xi$ as follows:
\begin{eqnarray}
\lefteqn{
    \!\int_0^1\!\! \frac{d\xi}{\xi} \,
    \!\int\!\!dP\!S_3(eq(p_2)\to\nu H^0 q)
    \times \int\!\!d^3k \, \delta^{(3)}(k - \frac{1-\xi}{\xi}p_2)
    \, \theta(1-\xi-\delta)
}
\nn
& = &
    \!\int_0^1\!\! \frac{d\xi}{\xi} \,
    \int\!\! dP\!S_4(eq(p_3)\to\nu H^0 q g)
    (2\pi)^3 2k_0\,\delta^{(3)}(k - (1-\xi)p_3)
    \, \theta(1-\xi-\delta)
\nn
& = &
    \int\!\! d\hat{s}
    \int\!\! dP\!S_3(eq(p_3)\to\nu H^0 q^*)
    \times
    \!\int_0^1\!\! \frac{d\xi}{\xi} \,
    \int\!\!dP\!S_2(q^* \to qg)\,(2\pi)^2
    2k_0\,\delta^{(3)}(k - (1-\xi)p_3)
\nn&&\quad\quad\quad\times
    \, \theta(1-\xi-\delta)
\nn
& = &
    \int\!\! d\hat{s}
    \int\!\! dP\!S_3(eq(p_3)\to\nu H^0 q^*) \, \frac{1}{Q_2^2}
    \,\theta(\hat{s}-\Delta\frac{Q_2^2}{Q_2^2-\Delta})
\end{eqnarray}
where $p_2=xP$ and $p_3=p_2+k=zP$, and terms of $\order(m^2/Q_2^2)$ are
neglected.
The $2\to3$ phase space integrals in the hard Bremsstrahlung and counter term
now have the same
kinematical boundaries. This means that the subtraction can be performed
for each point in this phase space.  The
integrals over $\hat{s}$ are also seen to be
identical in the limit $\Delta\to0$.
What remains are integrals over
$\cos\theta$ and $\phi$ for the Bremsstrahlung term ($\cos\theta,\phi$ are
the angle variables in the $q^*$ cms), which do not have a counterpart in
the counter term.

In the integration of the hard and glue contributions, eqs \ref{eq:hardquark}
and \ref{eq:matglue}, it will be necessary to treat the single and double pole
terms separately.  The double pole terms are non-divergent (the numerator is
proportional to $m^2$), but give a finite contribution.  They have a malicious
$\cos\theta$ dependence, but can be integrated numerically with a suitable
mapping.

In the rest, the integral over $\cos\theta$ gives an exact
cancellation of the single pole terms in the collinear limit $\cos\theta\to1$
when the radiative terms are symmetrized in $\cos\theta$, and the collinear
logarithm in the counter term (see eqs (\ref{eq:Pqq}) and (\ref{eq:PqG}))
is represented by
\begin{equation}
\ln(\xi(1-\xi)^n m^2/Q_2^2) = \int_0^1\!\!
    \frac{d\,\cos\theta}{\cos\theta - (1-\xi(1-\xi)^n m^2/Q_2^2)^{-1}}
\end{equation}
with $n = \pm1$ for the quark and gluon contributions, respectively. Now the
sum of the radiative contribution, excluding the double pole terms, and the
counterterms is finite for all values of $\cos\theta$.  As the boundaries on
all integrals coincide this sum can be integrated numerically.

For clarity we repeat that
as the virtual and soft contributions were seen to cancel against their
respective counter terms, only the above integrals, which combine the hard
Bremsstrahlung and gluonic contribution with their respective counter terms,
have to be evaluated numerically.
The independence of the result on the soft
cut-off $\Delta$ and the quark mass $m$ was verified by varying the value used
in the computation.  The numerical accuracy now
suffices: without cuts we find agreement with the analytical calculation
presented in the previous section within the statistical accuracy (1--3\%).

When applying the cuts one has to define the way to combine the four jets
produced in $eq \to \nu H^0 qg$ and $eg \to \nu H^0 q \bar{q}$ with $H^0 \to
b\bar{b}$ into three jets.  The algorithm used
was to combine the two partons with minimal distance in the $\eta$--$\phi$
plane if this distance was less than 1, and else to require that one of the
partons did not pass the $(p_\perp,\eta)$ cuts. The resulting 3-jet event
was then subjected to the cuts described in section~\ref{sec-born}.  In the
collinear and soft limits this is seen to be equivalent to the cuts applied to
the Born approximation.\footnote{The same algorithm was used in the
calculation presented in \cite{BORN3} to estimate the contribution to the
background from four jet events.}

With the experimental cuts applied a numerical accuracy of about 10\% of the
corrections is achieved.  The reduction in accuracy is
caused by small regions in phase space,
close to the collinear limit $\cos\theta=1$, where the three jet signal is cut
but the four jet survives, or vice versa. Of course {\em at\/} the collinear
limit the kinematical cuts are the same, but for any finite angle between the
quark and gluon they differ.  The resulting wedge-shaped regions have the
original logarithmic singularity, and hence give a finite but small
contribution.Unfortunately present-day numerical integration methods
\cite{VEGAS} do not handle such highly discontinuous functions in a
high-dimensional phase space very well.

The results of this calculation are summarized in fig.~\ref{F7}.  One can see
that the effect of the cuts is to make the corrections negative and larger.
This is to be expected, as the requirement to observe exactly three jets cuts
away part of the (positive) Bremsstrahlung terms.  The almost exact
cancellation which keeps the integrated corrections so small is now violated,
thus are the relative corrections larger with cuts taken into consideration.
The gluonic contribution is much smaller as the gluonic component tends to be
softer, which reduces the probability that four jets are observed.

\subsubsection{Corrections to the decay}
\label{sec-QCD22}

The $\order(\alpha_s)$ correction $\Gamma^{(1)}$ to the decay width of the
Higgs boson into $b$-quarks has long been known \cite{BRANCH2}.  The diagrams
for this correction are shown in fig.~\ref{F8}. Recently also the
$\order(\alpha_s^2)$ contributions $\Gamma^{(2)}$ were evaluated
\cite{BRANCH3}.  In the mass range $80 \leq M_H \leq 150 \GeV$ and adopting
the \MSbar\ scheme, the relative corrections $\Gamma^{(1+2)}/\Gamma^{(0)}$
vary from $-28 \%$ to $-43 \%$. In the applications considered so far in this
paper only the branching ratio $Br(H^0 \rightarrow b\bar{b})$ enters.  This is
affected less than the $b\bar{b}$ partial width, at least as long as $H^0 \to
b\bar{b}$ is the dominant decay mode. Numerically, we derive from
ref.~\cite{BRANCH1A} that $-3.4\% \gsim Br^{(1+2)}/Br^{(0)} \gsim -36\%$ in the
mass range considered.\footnote{The QED and electroweak corrections to
$\Gamma(H^0 \rightarrow b\bar{b})$ were found in ref.~\cite{BAR} to vary
between  $-0.6\%$ and $-1.7\%$ for $50 < M_H < 150 \GeV$.  They have not been
included here.}

However, these corrections also change in the presence of kinematical cuts.
For the moment, let us consider the $\order(\alpha_s)$ corrections only.  A
part of these corrections involves the emission of a hard gluon, $H^0 \to
b\bar{b}g$, which can change the configuration of the bottom jets
considerably. These events should not all be included in the observable cross
section as we demand exactly three jets, of which two originate in the bottom
quarks.  The difference must be considered an $\order(\alpha_s)$ correction
to the results given in section \ref{sec-born}, which already
include the QCD corrected branching ratio $Br(H^0\to b\bar{b})$.  This
correction is scheme independent as only the hard radiation part is
involved.  It is also expected to be relatively small, as
the main effect of the integrated corrections is to introduce a running $b$
quark mass in the Yukawa coupling.  These terms are of order
$\ln(M_H^2/m_b^2)$; they arise from the renormalisation of the $b\bar{b}H$
coupling and do not contribute here.  Furthermore, the effect of the cuts
clearly makes the $\order(\alpha_s)$ contribution negative.

The $\order(\alpha_s)$ correction to $\sigma^{(0)}_{obs}$ described above is
given by the difference
\begin{equation}
\label{eq:difsig}
    \sigma^{(1)}_{decay} = \sigma^{(1)}_{b\bar{b}g} - \sigma^{(1)}_{b\bar{b}}
\end{equation}
where
\begin{eqnarray}
    \sigma^{(1)}_{b\bar{b}g} & = & \int\!\! dx
        \int\!\! dP\!S_3(eq\to \nu H^0 q)
        \sum_{i=1}^2 {|\mat_i^{(0)}|^2} q_i(x,Q_2^2)
\nn&&\mbox\quad\quad
        \times { \frac{1}{2M_H}
        \int\!\! dP\!S_3(H^0\to b\bar{b}g)
        {|\mat_{H^0\to b\bar{b}g}|^2}
        \theta(\mbox{4-jet cuts})
        } \bigm/ {\Gamma_{tot}}
\end{eqnarray}
is the contribution from hard gluon radiation subject to the appropriate 4-jet
cuts, while
\begin{eqnarray}
    \sigma^{(1)}_{b\bar{b}} & = & \int\!\! dx
        \int\!\! dP\!S_3(eq\to \nu H^0 q)
        \sum_{i=1}^2 {|\mat_i^{(0)}|^2} q_i(x,Q_2^2)
        \frac{\int\!\! dP\!S_2(H^0\to b\bar{b}) \theta(\mbox{3-jet cuts})}
            {\int\!\! dP\!S_2(H^0\to b\bar{b})}
\nn&&\mbox{}\quad\quad\times
        { \frac{1}{2M_H} \int\!\! dP\!S_3(H^0\to b\bar{b}g)
        {|\mat_{H^0\to b\bar{b}g}|^2}} \bigm/ {\Gamma_{tot}}
\end{eqnarray}
is the corresponding integrated contribution corrected for the effect of the
3-jet cuts.  $\Gamma_{tot}$ is the total width of the Higgs boson.  Obviously
in the absence of cuts $\sigma^{(1)}_{decay} = 0$.

The decay matrix element, summed over all spins, ${|\mat_{H^0\to
b\bar{b}g}|^2}$, does not depend on the orientation of the $b$ quark in the
Higgs boson cms system; only the 4-jet cuts will be influenced by it.  We thus
extract the integral over these angles $\int d\Omega_b$ from the 3-body decay
phase space integral.  The 2-body phase space integral contains a similar
angle, which enters only in the 3-jet cuts.  Identifying these two angles we
obtain
\begin{eqnarray}
    \sigma^{(1)}_{decay} & = & \int\!\! dx
    \int\!\! dP\!S_3(eq\to \nu H^0 q)
        \sum_{i=1}^2 {|\mat_i^{(0)}|^2} q_i(x,Q_2^2)
\nn&&\mbox{}\quad\quad\times
        { \frac{1}{2M_H}
        \frac{1}{(2\pi)^5} \frac{1}{8} \int\!\!d\phi\,dE_b\,dE_{\bar{b}}
        {|\mat_{H^0\to b\bar{b}g}|^2}} \bigm/ {\Gamma_{tot}}
\nn&&\mbox{}\quad\quad
    \times \int\!\! d\Omega_b
        \left\{ \theta(\mbox{4-jet cuts}) - \theta(\mbox{3-jet cuts})
        \right\}.  
\end{eqnarray}
where $\phi$ is the azimuthal angle of the $\bar{b}$ momentum with respect to
the $b$ momentum.
In a Monte Carlo integration this means that for each Bremsstrahlung event one
constructs a decay into only $b\bar{b}$ with the $b$ quark momentum in the
same direction (in the Higgs boson cms).  One examines whether this
alternative kinematical configuration passes the 3-jet cuts: the non-zero
region is the one in which the 3-jet configuration passes its cuts,  but the
4-jet does not, or vice versa.  This will not happen in the limit of soft or
collinear gluon emission, so the result is free of singularities and
$\ln(M_H^2/m_b^2)$ terms.

Finally, we can include important $\order(\alpha_s^2)$ effects by
reintroducing the 2-loop integrated branching ratio as
\begin{equation}
\label{eq:reintroduce}
    \frac{{|\mat_{H^0\to b\bar{b}g}|^2}}{\Gamma_{tot}} =
    \frac{{|\mat_{H^0\to b\bar{b}g}|^2}}{\Gamma^{(0+1+2)}}
    Br^{(0+1+2)}.
\end{equation}
The fraction on the r.h.s.\ of eq.~\ref{eq:reintroduce} is now almost
independent of the (running) $b$ quark mass, the main effect of which has been
absorbed in the branching ratio.  Effectively, we thus use the two loop
expression for the $b\bar{b}H^0$ coupling (which is proportional to the $b$
quark mass) everywhere, while considering the effects of kinematical cuts only
up to $\order(\alpha_s)$.

The expression for ${|\mat_{H^0\to b\bar{b}g}|^2}$ used agrees with previous
calculations \cite{BRANCH2}.  It is given in the appendix (eq.
(\ref{eq:matdecay})).  The resulting correction $\sigma^{(1)}_{decay}$ is
shown in fig.~\ref{F7}. The numerical result is not sensitive to the precise
value of $m_b$ in the range $3\leq m_b \leq 8\GeV$.  The correction
$\sigma^{(1)}_{decay}$ to the decay is larger than the correction
$\sigma^{(1)}_{prod}$ to the production process as the probability to detect a
fourth jet radiated from a final state $b$-quark is higher.  The rise for low
Higgs mass results from the high probability that the $b$ and $\bar{b}$ jet
become indistinguishable when a hard gluon is emitted in a boosted system.

It should be noted that the size of these corrections critically depends on
the way in which the jets are reconstructed.  Here we employed a rather simple
method, which could be improved upon by using more sophisticated
algorithms.  The Monte Carlo approach given here allows such studies to be
performed in a straightforward way.



\section{QED Corrections}
\label{sec-QED}
The dominant contributions to the QED corrections for $ep$ scattering at high
energies can be described in the leading logarithmic approximation
\cite{RC1,RC2}. For charged current processes, only initial state
bremsstrahlung (see fig.~\ref{F9}) from the incoming electron line has to be
considered, since radiation from the quark line merely leads to a small extra
contribution of $\order(3 \alpha e_q^2 /4 \alpha_s)$ to the QCD evolution.  In
this approximation, the QED correction to the Higgs production cross section
is given by
\begin{equation}
\label{QEDSI}
\sigma^{(1)}(QED) =  \frac{\alpha}{2 \pi} \int\!\! dz
    \frac{1+z^2}{1-z} \ln \left( \frac{\mu^2} {m^2_e} \right)  \left\{
    \sigma^{(0)}(zp_1) - \sigma^{(0)}(p_1) \right\}
\end{equation}
where $\alpha = 1/137$.  The
factorization scale $\mu$ is related to the $k_{\perp}-$integral of the
radiated photon, $\mu^2 \approx Q_1^2 \approx s$.  We have chosen $\mu^2=s$
(see ref.\ \cite{RC1}).

In fig.~\ref{F10} the QED correction $\sigma^{(1)}(QED)$ to the total
production cross section is shown as a function of $M_H$ using the
parametrization MT (DIS, set 2) \cite{MT}.  In the mass range considered,
it amounts to
$-4\%$ to $-5\%$ for the choice $\mu^2 =s$.
The QED radiative corrections are rather small because of
the small ratio $M_H^2/s \lsim 0.012$. A similar behaviour was recently
observed for the correction to heavy flavour production in neutral current
$ep-$scattering \cite{RC4}, where for small values of $4 m_Q^2/s$ only small
negative corrections are obtained.  The correction to the observable cross
section after cuts is also shown in fig.~\ref{F10}. The relative correction
$\sigma^{(1)}_{obs}(QED)/\sigma^{(0)}_{obs} = -4.5\%$ to
$-5.8\%$.\footnote{The leading QED corrections to the $W$ and top backgrounds
were also calculated and found to be comparable to the corrections to the
signal.} These results are only weakly influenced by the choice of the parton
distributions.


\section{Conclusions}
\label{sec-concl}

We calculated the $\order(\alpha_s)$ QCD corrections to Higgs boson production
in $e^-p$ collisions via the $WW$ fusion process followed by the decay $H^0
\rightarrow b\bar{b}$ in the zero-width approximation. Results are
obtained for the fully integrated cross section, as well as for the observable
cross section after application of the kinematical cuts suggested in
ref.~\cite{BORN3}. We find surprisingly small corrections of the order of a
percent and less for the uncutted cross section, and larger and negative
corrections when the cuts are included. Our results are at variance with the
expectation expressed in ref.~\cite{BORN3}.

We conclude that the QCD corrections that have to be added, once the known
corrections to the branching ratio are included, in order to obtain the
complete
$\order(\alpha_s)$ corrections to $e^-p \rightarrow \nu H^0 X; H^0
\rightarrow b\bar{b}$, are in fact very small; the only sizeable effects
originate in the selection cuts which mainly affect the hard gluon radiation.
The following factors contribute to make the corrections this small:
\begin{itemize}
\item The cross section is dominated by $F_2$, which is not renormalized in the
DIS scheme.  The difference between the DIS and \MSbar\ structure functions
suggest that the corrections would change by a few percent when changing to the
\MSbar\ scheme.
\item The scale $Q_2^2$ is large everywhere ($Q_2^2 \gsim 100 \GeV^2$).
This means that $\alpha_s$ is small ($\alpha_s/2\pi \approx 0.02$).
\item No large logarithms enhance the corrections, as these have
already been absorbed into the evolution of the distribution
functions.
\end{itemize}
The main effect due to the cuts occurs at the decay vertex,
where the extra gluon is either
detected as a fourth jet or causes the $b$ jets to be too close together.
These effects however depend on the algorithms used to define the jets.

In addition, we also calculated the effect due to initial state QED
bremsstrahlung from the electron. This correction decreases the observable
cross section by roughly $5\%$.  The relative QCD and QED corrections to  the
observable cross section are summarized in fig.~\ref{F11}.  One can conclude
{}from our results that the analysis performed in ref.~\cite{BORN3} is not
changed significantly by higher order effects as far as the signal is
concerned. What remains to be done is a corresponding calculation for the
background processes.

\paragraph{Acknowledgement.}

We would like to thank J.~A.~M. Vermaseren, K.~J.~F. Gaemers and D.~Zeppenfeld
for discussions.


\newpage
\appendix
\setcounter{equation}{0}
\def\theequation{A.\arabic{equation}}
\section{Appendix}
\label{sec-append}

In this appendix we list explicit expressions used in the calculation of
the $\order(\alpha_s)$ QCD corrections to the observable cross section
as discussed in section \ref{sec-QCD2}.

In the regularization scheme adopted in this calculation, the
$\order(\alpha_s)$ modifications of the quark distributions due to the
virtual, soft and hard Bremsstrahlung terms and the gluon contribution are
given by
\begin{eqnarray}
\label{eq:countervir}
q^{(1)}_{i,virt}(x,Q^2) & = & \frac{\alpha_s}{2\pi} C_F\Bigl\{ - 4 +
    \frac{\pi^2}{3} + \ln\Bigl(\frac{m^2Q^2}{\lambda^4}\Bigr)
    \ln\Bigl(\frac{m^2}{Q^2}\Bigr) - 2\ln\Bigl(\frac{\lambda^2}{m^2}\Bigr)
    - 3\ln\Bigl(\frac{m^2}{Q^2}\Bigr) \Bigr\} q_i(x,Q^2)
\nn
\\
q^{(1)}_{i,soft}(x,Q^2) & = & \frac{\alpha_s}{2\pi} C_F \Bigl\{
    + \frac{3}{2} - \frac{2\pi^2}{3} + \frac{1}{2}\ln\Bigl(\frac{\Delta}
    {m^2}\Bigr) + 2\ln\Bigl(\frac{\lambda^2Q^2}{\Delta^2}\Bigr)
    \biggl( 1 + \ln\Bigl(\frac{m^2}{Q^2}\Bigr) \biggr) -
    \ln^2\Bigl(\frac{\Delta}{m^2}\Bigr) \Bigr\}
\nn&&\quad\quad\times q_i(x,Q^2)
\\
q^{(1)}_{i,hard}(x,Q^2) & = & \frac{\alpha_s}{2\pi}
\int_x^{1-\delta}\frac{d\xi}{\xi}
\widehat{P}_{qq}\Bigl(\xi,\frac{m^2}{Q^2}\Bigr) q_i(x/\xi,Q^2)
\\
\label{eq:counterglue}
q^{(1)}_{i,glue}(x,Q^2) & = & \frac{\alpha_s}{2\pi}
\int_x^1\frac{d\xi}{\xi}
\widehat{P}_{qG}\Bigl(\xi,\frac{m^2}{Q^2}\Bigr) G(x/\xi,Q^2)
\end{eqnarray}
with $\Delta = (p_2'+k)^2_{min}$, $\delta = (\Delta-m^2)/Q^2$ and
\begin{eqnarray}
\label{eq:Pqq}
    \widehat{P}_{qq} \Bigl(\xi, \frac{m^2}{Q^2} \Bigr) &=& C_F\Biggl\{
        \frac{1+\xi^2}{1-\xi}\ln\Bigl[\frac{Q^2}{m^2\xi(1-\xi)}\Bigr] -
        \frac{3\xi^2+\xi-\frac{1}{2}}{1-\xi}\Biggr\}
\\
\label{eq:PqG}
    \widehat{P}_{qG} \Bigl(\xi, \frac{m^2}{Q^2} \Bigr) &=& \frac{1}{4}
        \Biggl\{ (1-2\xi+2\xi^2)\ln\Bigl[\frac{Q^2(1-\xi)}{m^2\xi}
        \Bigr] - 8\xi^2 + 8\xi - 1 \Biggr\}.
\end{eqnarray}
The dependence on $\lambda$ cancels in
the sum of the soft and virtual parts:
\begin{eqnarray}
\label{eq:qvirtsoft}
    q^{(1)}_{i,virt+soft}(x,Q^2) &=& \!- \frac{\alpha_s}{2\pi} C_F
    \left \{
    \frac{5}{2} + \frac{\pi^2}{3} + \frac{7}{2}\ln\Bigl(\frac{\Delta}{Q^2}
        \Bigr)
    + \ln^2\Bigl(\frac{\Delta}{Q^2}\Bigr)
    + 2 \ln\Bigl(\frac{m^2}{Q^2}\Bigr) \Bigl [
    \ln\Bigl(\frac{\Delta}{Q^2}\Bigr) + \frac{3}{4} \Bigr ]  \right \}
\nn&&\quad\quad
    \times q_i(x,Q^2).
\end{eqnarray}
Furthermore, $q^{(1)}_{i,hard}(x,Q^2)$ may be rewritten as
\begin{eqnarray}
    q^{(1)}_{i,hard}(x,Q^2) &=& \mbox{}
    \frac{\alpha_s}{2\pi} \int_0^{1 - \delta}  d\xi
    \left\{
    \widehat{P}_{qq} \Bigl(\xi, \frac{m^2}{Q^2} \Bigr)
    \Bigl[\theta(\xi - x)\frac{1}{\xi}q_i(x/\xi,Q^2) - q_i(x,Q^2)\Bigr]
    \right\}
\nn&&\mbox{}
    + \frac{\alpha_s}{2\pi} \int_0^{1 - \delta} d\xi
    \widehat{P}_{qq} \Bigl(\xi, \frac{m^2}{Q^2} \Bigr)
    q_i(x,Q^2).
\end{eqnarray}
In the first integral the limit $\delta \rightarrow 0$ can be taken because
terms of $\order(\delta)$ may be neglected.  The second integral can be
calculated analytically yielding
\begin{equation}
\label{eqds}
\frac{\alpha_s}{2\pi} C_F \left \{
\frac{5}{2} + \frac{\pi^2}{3} + \frac{7}{2} \ln\delta
+ \ln^2\delta
+ 2 \ln \left ( \frac{m^2}{Q^2} \right ) \left [ \ln\delta
+ \frac{3}{4} \right ] \right \} q_i(x,Q^2).
\end{equation}
In the limit $m\to0$, the contribution (\ref{eqds}) just cancels the term
(\ref{eq:qvirtsoft}) in the sum $q_{i,virt+soft} + q_{i,hard}$.

The differential cross sections for hard gluon Bremsstrahlung for an incoming
quark or antiquark and  the gluon contribution were derived using FORM
\cite{FORM} and are given by
\begin{eqnarray}
\label{eq:hardquark}
d\sigma_{hard}^{(1)} & = &
    \frac{1}{2zs} \sum_{i=1}^2 |\mat_{i,hard}^{(1)}|^2q_i(z,Q_2^2)
    \,\theta(\hat{s}-\Delta)\,dP\!S_4\,dz
\nn& = &
    \frac{1}{2zs}
    \frac{g^6 M_W^2}{(Q_1^2 + M_W^2)^2 (Q_2^2 + M_W^2)^2}
    4\pi\alpha_s C_F \sum_{i=1}^2 \Biggl\{ \bigl[
       - (k p_1)(p_1' p_2')(p_3 p_2')/(k p_3)/(k p_2')
\nn&&\mbox{}
       + (k p_1)(p_1' p_2')/(k p_3)
       + (k p_1')(p_1 p_3)(p_3 p_2')/(k p_3)/(k p_2')
       + (k p_1')(p_1 p_3)/(k p_2')
\nn&&\mbox{}
       - (p_1 p_3)(p_1' p_3)/(k p_3)
       +2(p_1 p_3)(p_1' p_2')(p_3 p_2')/(k p_3)/(k p_2')
       - (p_1 p_3)(p_1' p_2')/(k p_3)
\nn&&\mbox{}
       + (p_1 p_3)(p_1' p_2')/(k p_2')
       + (p_1 p_2')(p_1' p_2')/(k p_2')
       + m^2(k p_1)(p_1' p_2')/(k p_3)^2
\nn&&\mbox{}
       - m^2(k p_1')(p_1 p_3)/(k p_2')^2
       - m^2(p_1 p_3)(p_1' p_2')/(k p_3)^2
       - m^2(p_1 p_3)(p_1' p_2')/(k p_2')^2
    \bigr]
\nn&&\mbox{}
    \times u_i(z,Q_2^2)
       + \bigl[ p_1 \leftrightarrow p_1' \bigr] \bar{d}_i(z,Q^2) \Biggr\}
       \,\theta(\hat{s}-\Delta)\, dP\!S_4(eq\to\nu H^0qg)\,dz,
\\
\label{eq:matglue}
d\sigma_{glue}^{(1)} & = &
    \frac{1}{2zs} |\mat_{glue}^{(1)}|^2 G(z,Q_2^2)
    \,dP\!S_4\,dz
\nn& = &
    \frac{1}{2zs}
    \frac{g^6 M_W^2}{(Q_1^2 + M_W^2)^2 (Q_2^2 + M_W^2)^2}
    4\pi\alpha_s \frac{1}{2} 2 \Biggl\{
       + (k p_1)(k p_1')/(k p_3)
\nn&&\mbox{}
       +2(k p_1)(k p_2')(p_1' p_2')/(k p_3)/(p_3 p_2')
       - (k p_1)(k p_2')(p_1' p_3)/(k p_3)/(p_3 p_2')
\nn&&\mbox{}
       - (k p_1)(p_1' p_2')/(k p_3)
       - (k p_1)(p_1' p_2')/(p_3 p_2')
       + (k p_1)(p_1' p_3)/(p_3 p_2')
\nn&&\mbox{}
       - (k p_2')(p_1 p_3)(p_1' p_2')/(k p_3)/(p_3 p_2')
       + (p_1 p_3)(p_1' p_2')/(k p_3)
       + (p_1 p_2')(p_1' p_2')/(p_3 p_2')
\nn&&\mbox{}
       + m^2(k p_1)(p_1' p_3)/(p_3 p_2')^2
       - m^2(k p_1)(p_1' p_2')/(k p_3)^2
       - m^2(k p_1)(p_1' p_2')/(p_3 p_2')^2
\nn&&\mbox{}
       + m^2(p_1 p_3)(p_1' p_2')/(k p_3)^2
    \Biggr\} G(z,Q_2^2)\, dP\!S_4(eg\to\nu H^0q\bar{q})\,dz.
\end{eqnarray}
Here $k$ denotes the  momentum of the extra outgoing parton and  $p_3 = zP$ is
the momentum of the incoming parton.  The decay $H^0\to b\bar{b}$ is again
inserted when the observable cross section is calculated.

The matrix element squared for the hard Bremsstrahlung part of the decay width
used in section \ref{sec-QCD22} is given by
\begin{eqnarray}
\label{eq:matdecay}
    |\mat_{H^0\to b\bar{b}g}|^2 & = & \frac{3g^2}{32} \frac{m_b^2}{M_W^2}
    \; 4\pi\alpha_s C_F \Bigl\{
       + 8
       + 4 (p_b k) /(p_{\bar{b}} k)
       + 8 (p_b p_{\bar{b}}) /(p_b k)
       + 8 (p_b p_{\bar{b}}) /(p_{\bar{b}} k)
\nonumber\\
&&\hspace*{-1.1cm}\mbox{}
       + 8 (p_b p_{\bar{b}})^2 /(p_b k) /(p_{\bar{b}} k)
       + 4 (p_{\bar{b}} k) /(p_b k)
       - 4 m_b^2 (p_b k) /(p_{\bar{b}} k)^2
       - 4 m_b^2 (p_b p_{\bar{b}}) /(p_b k)^2
\nonumber\\
&&\hspace*{-1.1cm}\mbox{}
       - 4 m_b^2 (p_b p_{\bar{b}}) /(p_{\bar{b}} k)^2
       - 8 m_b^2 (p_b p_{\bar{b}}) /(p_b k) /(p_{\bar{b}} k)
       - 4 m_b^2 /(p_b k)
       - 4 m_b^2 /(p_b k)^2 (p_{\bar{b}} k)
\nonumber\\
&&\hspace*{-1.1cm}\mbox{}
       - 4 m_b^2 /(p_{\bar{b}} k)
       + 4 m_b^4 /(p_b k)^2
       + 4 m_b^4 /(p_{\bar{b}} k)^2 \Bigr\}.
\end{eqnarray}
The two-loop corrected decay width for $H^0 \to b\bar{b}$ is derived in
ref.\ \cite{BRANCH2}:
\begin{equation}
   \Gamma^{(0+1+2)} = \frac{3g^2}{32\pi} \frac{\tilde{m_b}^2}{M_W^2}
        \frac{\bigl( M_H^2-4\tilde{m_b}^2 \bigr)^{3/2}}{M_H^2}
        \,(1+1.803\alpha_s+2.953\alpha_s^2),
\end{equation}
with the running $b$ quark mass $\tilde{m_b}$ and coupling constant $\alpha_s$
taken at the scale $M_H^2$.  As stated
in the text, the magnitude of the $\order(\alpha_s)$ corrections discussed in
this paper does not depend significantly on the value chosen for this mass.


\section{Figures}

\begin{figure}[hbtp]
\caption{The Born diagram for Higgs boson production via $WW$ fusion.}
\label{F1}
\end{figure}

\begin{figure}[hbtp]
\caption[Born cross section]
{Born cross sections using the parton
distributions by Morfin and Tung (MT2). Long-dashed line:
$\sigma^{(0)}_{tot}$
for MT2(\MSbar); full line: $\sigma^{(0)}_{tot}$ for MT2(DIS);
dash-dotted line:
$\sigma^{(0)}_{tot} \times Br(H^0~\rightarrow~b\bar{b})$ for
MT2(DIS); short-dashed line: $\sigma^{(0)}_{obs}$  as defined in the text.}
\label{F2}
\end{figure}

\begin{figure}[hbtp]
\caption[Differential distributions in the Born approximation]
{Differential
distributions in $x$, $Q^2_2$, the transverse momentum
$p_\perp$ and
rapidity $\eta$ of the Higgs boson
for $M_H$ = 100 GeV using the Born approximation.
The dashed lines give the
distributions after application of the selection cuts proposed in ref.\
\cite{BORN3}.}
\label{F3}
\end{figure}

\begin{figure}[hbtp]
\caption{The $\order(\alpha_s)$ diagrams contributing to the QCD corrections
at the hadronic vertex.}
\label{F5}
\end{figure}

\begin{figure}[btp]
\caption[The $\order(\alpha_s)$ QCD corrections to $\sigma_{tot}$]{The
$\order(\alpha_s)$ QCD corrections to $\sigma_{tot}$ in the DIS scheme. Full
line:  $\sigma^{(1)}_{tot}(QCD)$, long-dashed line: $\sigma_L^q$,
     dash-dotted
line: $\sigma_L^G$, and short-dashed line:\ $\sigma_3^q$.}
\label{F6}
\end{figure}

\begin{figure}[hbtp]
\caption[The $\order(\alpha_s)$ QCD corrections for the cuts]
{The $\order(\alpha_s)$ QCD corrections  including the cuts
explained in section~\ref{sec-born}.
The corrections to the production process $ep\to\nu H^0X$ from quarks
and gluons and to the decay process $H^0\to b\bar{b}$ are shown
separately.
The total sum $\sigma^{(1)}_{obs}(QCD)$
is the appropriate $\order(\alpha_s)$ correction to the observable cross
section $\sigma^{(0)}_{obs}$ (which
already includes the QCD corrected $Br(H^0 \rightarrow b\bar{b})$).
The error bars indicate the
precision of the Monte Carlo calculation.}
\label{F7}
\end{figure}

\begin{figure}[hbtp]
\caption{The $\order(\alpha_s)$ diagrams contributing
to the QCD corrections to the decay vertex.}
\label{F8}
\end{figure}

\begin{figure}[hbtp]
\caption{Diagram representing the leading QED correction
to charged current $ep$ processes.}
\label{F9}
\end{figure}

\begin{figure}[btp]
\caption[The leading QED corrections]
{The leading QED corrections to
the total (dashed line) and observable (full line) cross sections
for $e^-p \rightarrow \nu H^0 X; H^0 \rightarrow b\bar{b}$ (using $\mu^2=s$).}
\label{F10}
\end{figure}

\begin{figure}[btp]
\caption[The relative corrections]
{
The relative corrections
$\sigma^{(1)}_{obs}/\sigma^{(0)}_{obs}$
to the observable
cross section for $e^-p\to \nu H^0 X; H^0\to b\bar{b}$, including
the cuts
given in section~\ref{sec-born}.
The QCD and the leading QED
corrections are shown separately. The
error bars indicate the statistical precision of the Monte Carlo
calculation.
}
\label{F11}
\end{figure}
\mbox{}

\newpage
\begin{thebibliography}{99}

\bibitem{PPCOL}
J.~F. Gunion, H.E. Haber, G.~L. Kane and S. Dawson, The Higgs
Hunter's Guide, Addison-Wesley, 1990;  \\
G. Altarelli,
Proc.\  of the ECFA Large Hadron Collider Workshop, Aachen,
1990, Eds.\  G.~Jarlskog, D.~Rein, CERN 90--10,
ECFA 90--133, Vol.\  {\bf 1}, p. 153;\\
D. Denegri, ibid.,\ Vol.\  {\bf 1}, p. 56.
\bibitem{XX}
W. Marciano and F. Page, Phys. Rev. Lett. {\bf 66} (1991) 2433;\\
J.F. Gunion, Phys. Lett. {\bf B261} (1991) 510.
\bibitem{JLC}
see e.g.\ Proc.\ of the 1st Japanese Linear Collider (JLC) Workshop,
1989, Ed. S. Kawabata, KEK, Tsukuba, 1989;\\
Proc.\ of the Workshop `$e^+e^-$ Collisions at 500 GeV', Ed.\ P.\ Zerwas,
DESY, Hamburg, 1992 (in press).
\bibitem{LUM}
G. Brianti,
Proc.\  of the ECFA Large Hadron Collider Workshop, Aachen,
1990, Eds.\  G.~Jarlskog, D.~Rein, CERN 90--10,
ECFA 90--133, Vol.\  {\bf 1}, p. 40;\\
W. Bartel, ibid., Vol.\  {\bf 3}, p. 824;\\
A. Verdier, CERN--SL--90--89--AP.
\bibitem{BORN1}
S. Midorikawa and M. Yoshimura, Nucl.\  Phys.\
{\bf B162} (1980) 365; \\
Z. Hioki, S. Midorikawa and H. Nishiura, Progr.\  Theor.\  Phys.\
{\bf 69} (1983) 1484;\\
T. Han and H.~C. Liu, Z.\ Phys.\  {\bf C28} (1985) 295; \\
D.~A. Dicus and S.~S.~D. Willenbrock, Phys.\  Rev.\  {\bf D32}
(1985) 1642;\\
R. Bates and J.N. Ng, Phys.\  Rev.\  {\bf D33} (1986) 657; \\
K.~J.~F. Gaemers, R.~M. Godbole and M. van der Horst, Proc.\  of the
Workshop
'Physics at HERA', 1987, Ed.\  R.~D. Peccei, DESY, Hamburg,
1988, Vol.\  {\bf 2}, p.739.
\bibitem{BORN2}
G. Altarelli, B. Mele and F. Pitolli, Nucl.\  Phys.\
{\bf B287} (1987) 205.
\bibitem{BORN3}
D. Zeppenfeld, Proc.\  of the ECFA Large Hadron Collider Workshop,
Aachen,
 1990, Eds.\  G. Jarlskog, D. Rein, CERN 90--10,
ECFA 90--133, Vol.\  {\bf 1}, p. 268; \\
G. Grindhammer, D. Haidt, J. Ohnemus, J. Vermaseren and
D. Zeppenfeld,
ibid., Vol.\  {\bf 2}, p. 967.
\bibitem{DELPHI}
Ch. de la Vaissiere and  S. Palma--Lopes, DELPHI note 89--32 PHYS
38 (1989).
\bibitem{CCQCD}
J.~J. van der Bij and G.~J. van Oldenborgh, Z.\ Phys.\ {\bf C51} (1991) 477.
\bibitem{BRANCH1}
B. Kniehl, Phys.\  Lett.\  {\bf B244} (1990) 537.
\bibitem{BRANCH1A}
R. Kleiss, Z. Kunszt and W.~J. Stirling, Phys.\  Lett.\
{\bf B253} (1991) 269.
\bibitem{MT}
J. Morfin and Wu-Ki Tung, Z.\ Phys.\ {\bf C52} (1991) 13.
\bibitem{DIS}
G. Altarelli, J.~K. Ellis and G. Martinelli, Nucl.\  Phys.\
{\bf B157} (1979) 461.
\bibitem{HMRS}
P. Harriman, A. Martin, R. Roberts and J. Stirling, Phys.\
Rev.\  {\bf D42} (1990) 798.
\bibitem{DFLM}
M. Diemoz, F. Ferroni, E. Longo and G. Martinelli,
Z.\  Phys.\  {\bf C39} (1988) 21.
\bibitem{DO}
D. Duke and J.~F. Owens, Phys.\  Rev.\  {\bf D30} (1984) 49.
\bibitem{SCREEN}
L.~V. Gribov, E.~M. Levin and M.~G. Ryskin, Phys.\  Rep.\
{\bf 100} (1982) 1;\\
J. Bartels, J. Bl\"umlein and  G.~A. Schuler, Nucl.\
Phys.\  B (Proc.\  Suppl.) {\bf 18C} (1990) 147; \\
Z.\ Phys.\  {\bf C 50}  (1991) 91; \\
J. Collins and J. Kwiecinski, Nucl.\  Phys.\  {\bf B335}
(1990) 89.
\bibitem{KMRS}
J. Kwiecinski, A. Martin, R. Roberts and J. Stirling,
Phys.\  Rev.\
{\bf D42} (1990) 3645.
\bibitem{STRUC1}
G. Altarelli, R.~K. Ellis and G. Martinelli, Nucl. Phys. {\bf B143}
(1978) 521; {\bf B146} (1978) 544; {\bf B157} (1979) 461.
\bibitem{STRUC2}
B. Humpert and W.~L. van Neerven, Nucl.\  Phys.\
{\bf B184} (1981) 225.
\bibitem{STRUC3}
J.~G. K\"orner, E. Mirkes and G.~A. Schuler, Int.\
J.\  Mod.\  Phys.\
{\bf A4} (1989) 1781.
\bibitem{MASS}
J.~A.~M.~Vermaseren, K.~J.~F.~Gaemers and S.~J.~Oldham,
\newblock Nucl.~Phys. {\bf B187}  (1981) 301.\\
G.~J.~van Oldenborgh, {\em One-loop Calculations with
Heavy Particles},
Amsterdam Ph.~D. thesis, 1990.
\bibitem{VEGAS}
G.~P. Lepage, VEGAS, J.\  Comp.\ Phys.\ {\bf 27} (1978) 192.
\bibitem{BRANCH2}
E. Braaten and J.~P. Leveille, Phys.\  Rev.\  {\bf D22}  (1980) 715;\\
N. Sakai, Phys.\  Rev.\  {\bf D22}  (1980) 2220;\\
T. Inami and T. Kubota, Nucl.\  Phys.\  {\bf B179}  (1981) 171; \\
M. Drees and  Ken-ichi Hikasa, Phys.\  Lett.\ {\bf B240}
(1990) 445;\\
\
P. Kalyniak, N. Sinha, R. Sinha and J.~N. Ng, Phys.\ Rev.\ {\bf D43} (1991)
3664.
\
\bibitem{BRANCH3}
S.~G. Gorishny, A.~L. Kataev, S.~A. Larin and L.~R. Surguladze,
Mod.\  Phys.\  Lett.\ {\bf A5}  (1990) 2703; Phys.\  Rev.\
{\bf D43}  (1991) 1633.
\bibitem{BAR}
D.  Bardin, B.~M. Vilenskij and P.~C. Christova,
Sov. J. Nucl. Phys. {\bf 53}  (1991) 152;\\
B. Kniehl, Nucl.\ Phys.\ {\bf B376} (1992) 3.
\bibitem{RC1}
A. De Rujula, R. Petronzio and A. Savoy-Navarro, Nucl.\
Phys.\  {\bf B154}  (1979) 394; \\
M. Consoli and M. Greco, Nucl.\  Phys.\  {\bf B186}  (1981) 519.
\bibitem{RC2}
J. Bl\"umlein, Z.\  Phys.\  {\bf C47}  (1990) 89.
\bibitem{RC4}
J. Bl\"umlein, Phys.\ Lett.\ {\bf B271} (1991) 267.
\bibitem{FORM}
The symbolic manipulation program FORM was written by
J.~A.~M.~Vermaseren.
Version 1.0 of this program is available  via anonymous ftp from
nikhefh.nikhef.nl.
%
\end{thebibliography}
\end{document}